# Applying latent data assimilation to a fluid dynamics problem

Ruijia Yu

Imperial College London

ry122@ic.ac.uk

## Abstract

Shallow water equations are extensively considered in the domains of oceans, atmospheric modelling, and engineering research (Franca et al., 2022), which play significant roles in floods and tsunami governance. Nonetheless, the accurate prediction of shallow water behaviours is regarded as an arduous undertaking, particularly when confronted with multi-dimensional data and potential errors within the model. To address these challenges and improve the accuracy of forecasts, this study employs an integrated approach, involving dimensionality reduction methods, deep learning architectures, and data assimilation techniques. Indeed, Reduced-order modelling facilitates the conversions of high-dimensional data, extracting important features and attenuating the complexity of problems (Zhong et al., 2023). Subsequently, three different predictive models are utilized to prognosticate shallow water data in the reduced latent space, followed by comparisons of their prediction performance. Moreover, Bach and Ghil (2023) propose that through the amalgamation of model forecasts with observational metrics, the data assimilation algorithm can rectify their discrepancies, thereby enhancing the model's predictive prowess. Finally, the experimental results demonstrate that prediction values are congruent with actual observations, which accentuates the resilience and effectiveness of this comprehensive methodology. Its potential to accurately forecast shallow water data holds the applicability and referential significance in preventing storm swell and other meteorological events.

## 1. Introduction

Dellar and Salmon (2005) describe shallow water equations (SWE) as the simulations of incompressible fluid subjacent to the hydrostatic-balanced pressure surface. Their importance stems from the broad relevance and profound impact on handling water-related phenomena. In oceanography, SWE contributes to the prediction of tidal fluctuations, storm surges and wave propagation among coastal processes (Idier et al., 2019). Concurrently, SWE is instrumental in probing atmospheric currents, including the synergistic interactions between air and water bodies. For instance, Cavaleri and Rizzoli (1981) refer to the deployment of wind-wave forecasting models in the Northern Adriatic Sea to predict storms, with errors of effective wave height ranging from 10% to 20%. Moreover, there exists a Shallow Water

Axisymmetric Intensity Model, tailored for anticipating cyclonic phenomena characterized by transient intensity (Hendricks et al., 2021). Such prognostications enable scientists to make informed suggestions for disaster preparedness and sustainable development.

However, accurate SWE forecasts encounter several challenges. Firstly, an abundance of high-dimensional data should be processed, which involves diverse features, covering a broad temporal and spatial range. As posited by Han et al. (2012), increasing data dimensionality leads to a sparsity within the high-dimensional space, rendering the distance and similarity calculation arduous. Furthermore, equidistant samples exacerbate the difficulty of precisely distinguishing different categories (Sarkar et al., 2020), which hinders effective feature extraction. This "dimensional curse" induces a substantial rise in computational complexity and storage requirements, affecting the accuracy and efficiency of the anticipations (Jia et al., 2022). On the other hand, Ding et al. (2015) consider that excessive parameters in dealing with high-dimensional data could engender issues of overfitting, while insufficient samples may introduce some noise, attenuating the predictive generalization capability. Secondly, SWE comprises intricate physical and dynamic processes with nonlinear characteristics, such as the nonlinear propagation of wave height and topographical influences on water flow. Meanwhile, Pinto-Ledezma and Cavender-Bares (2021) point out the specific data like sea temperature, wind speed and topographic height, which are of diverse scales and distributions in SWE, presenting a rigorous test to the model's adaptability. Then, data uncertainty also constitutes some difficult obstacles in SWE predictions. There are certain measurement errors caused by environmental factors, impacting the accuracy of forecasts. For instance, rising sea levels and frequent extreme meteorological events under the auspices of climate change put forward higher requirements for prognostications (IPCC, 2019). Moreover, the observation data is usually constricted by climate variation, geological structures, and human intervention, leading to data deficiencies, outliers, and noise. Scheinost et al. (2019) explain their negative effect that the reliance on limited data may culminate in unstable anticipations. Simultaneously, according to Tripathi et al. (2021), the robustness of a model to noise and outliers may be compromised. Therefore, the continual refinement of accurate prognostic models can be regarded as an ongoing endeavour and meticulous attention to these difficulties is imperative to obtain precise forecasts.

Indeed, employing appropriate methods can effectively overcome the challenges. Currently, Reduced-order Modelling (ROM) emerges as a potent dimensionality reduction technology, especially suited for dealing with high-dimensional data and complex nonlinear relationships (Kadeethum et al., 2022). Moreover, Gong et al. (2022) indicate ROM helps decrease computational costs while upholding certain precision. As elucidated by Karasözen et al. (2020), ROM preserves SWE's quadratic structure through proper orthogonal decomposition and implicit methods for the temporal integration of partial differential equations. The primary objective of ROM is to transform high-dimensional data into latent variables that encapsulate the important patterns of SWE, effectively reducing parameters and dimensions to simplify the forecasting model (Kim et al., 2023). In addition, Hernandez et al. (2023) supplement that, within this latent space, each dimension corresponds to a significant feature, while the remaining dimensions tend to contain less information and noise. From this, ROM facilitates that the prediction model can focus on handling the most representative features, curtailing redundancy of information in data and augmenting the predictive accuracy.

Furthermore, data assimilation (DA) serves as a feasible approach to enhance anticipated performances, combining the forecasts with actual observed data (Cheng et al., 2023). Moye and Diekman (2018) contend DA can maximize the utility of data information, adjusting the model parameters and states, which is beneficial for alleviating errors and uncertainty of parameter estimation. Meanwhile, DA progressively optimizes model capability in long-term prognosticating and circumvents the accumulation of prediction errors (Zhuang et al., 2022).

For example, a seismic simulation applying ensemble DA foresees 70% of 21 large-scale quasi-periodic events, substantially outperforming periodic recursive models (Dinther et al., 2019) while Cheng et al. (2022) depend on latent DA techniques to improve the anticipated accuracy of wildfire with a reduction of 50% root-means-square-errors. Overall, the predictive model with DA adapts more expertly to real-world scenarios, diminishes some forecasting disparity, and attains more reliable outcomes.

This research adopts two different ROM methods, Principal Component Analysis (PCA), and Convolutional Autoencoder (CAE). PCA retains representative features with substantial variance, reflecting the principal axes of data variation (Rocchi et al., 2004). In contrast, CAE autonomously extracts features through convolutional and deconvolutional layers for the reconstructions. Subsequently, Long Short-Term Memory (LSTM), Random Forest (RF), and Polynomial Regression (PR) are employed to anticipate SWE behaviours within the latent space. LSTM is proficient in capturing long-term dependencies within SWE time series to predict temporally correlated phenomena like tides and flood levels (Le et al., 2019). Moreover, Khalaf et al. (2020) explicate that RF is employed for regression problems like water velocity and flood extent, while PR leverages polynomial features to fit complex relationships, suitable for nonlinear prediction. Finally, from a Bayesian perspective, combining machine learning with DA strengthens the models' explainability and robustness (Cheng et al., 2023). Thus, the Kalman Filter (KF) is deployed to update the estimations of model state accurately and timely with copious observational data.

This study's innovation is manifested in its pioneering application of latent DA in the field of computational fluid dynamics for the first time. In summary, it is articulated into three segments.

The first part is the application of CAE and PCA to map SWE data into latent space and compare their data reconstruction capabilities. After extracting 5000 data points from the SWE simulations, CAE and PCA models are constructed to achieve data dimension reduction and reconstruction. Their performance is then evaluated through metrics such as Mean Squared Error and R-squared values, along with reconstructed images to discern differences in feature extraction.

Next, three distinct predictive models, namely LSTM, RF, and PR, are utilized to forecast SWE behaviours in the latent space. Using PCA-processed "u" data from SWE as a specific object, the entire test set is input into these models for prediction without employing rollout. The comparison of predictive results with actual values in both latent and full spaces reveals the accuracy and stability of these three models in SWE data anticipations.

At last, we employ KF to ameliorate the prediction outcomes of three models. Based on the actual "u" data, KF can effectively rectify the prognosticated biases. Similarly, MSE and R-squared values as well as predictive curves, are displayed to facilitate the comparative result analysis after the implementation of KF.

## 2. Methodology

This section aims to elucidate the rationale behind the mentioned approaches in this study. With fundamental comprehension of these methods, we can enhance our grasp of their inherent characteristics and application scenarios, which are important for making effective strategies to address the current problem.

## 2.1 Convolutional Autoencoder (CAE)

Pintelas, E. and Pintelas, P. (2022) conduct that CAE is an unsupervised learning model based on convolutional neural networks for feature extraction and dimensional reduction. Its architecture consists of three fundamental components: Encoder, Decoder, and loss function. Represented as $z = E(x)$, the primary function of the Encoder involves compressing data $x$, which contains information from SWE images, into the latent space to obtain low-dimensional features. Subsequently, the Decoder takes charge of mapping $z$ back to the origin space, generating the reconstructed SWE images $x' = D(z)$. Moreover, network parameters are adjusted by the backpropagation concerned with the loss function minimization, which is the mean square error, $MSE = \frac{1}{n}\sum_{i=1}^{n}(x_i - x'_i)^2$ in this study.

There are typically multiple convolution layers and pooling layers in the Encoder. The former applies a filter that performs element-wise multiplication and traverses the entire data. This repetitive convolutional process facilitates the capture of patterns from the input SWE data. The calculation is shown, where $M, N$ are sizes of the filter $K$.

$$output\_image[i,j] = \sum_{m=0}^{M-1}\sum_{n=0}^{N-1} input\_image[i+m, j+n] * K[m,n]$$

Meanwhile, the purpose of pooling layers is to aggregate related information into single values and reduce the extracted features dimension. We specifically apply the Maximum pooling layer, which selects the largest feature value within non-overlapping regions $W$ as aggregate values (Zhai et al., 2017), to effectively preserve essential features of SWE data.

$$output\_image[i,j] = max_{(p,q) \in W} input\_image[p,q]$$

Furthermore, the Decoder module is composed of deconvolution layers and upsampling layers. Restoring the dimension of compacted features through padding is the responsibility of deconvolution layers while upsampling layers adopt bilinear interpolation to replicate information and allocate it to intended locations in the feature map.

$$output\_image[i,j] = \sum_{m=0}^{M-1}\sum_{n=0}^{N-1} input\_image[i-m, j-n] * K[m,n]$$

As for the model construction, we stack five layers in both Encoder and Decoder, which aids this CAE to extract features layer by layer and learn the higher-level feature representation.

## 2.2 Principal Component Analysis (PCA)

The conceptual origin of PCA can be traced back to the seminal works of Pearson (1901) and Hotelling (1933), which is widely employed for analysing extensive datasets characterized by numerous features. The principal objective is to reduce the dimensionality of multivariate data while maximizing the retention of pertinent information.

This process begins by computing the mean of the SWE dataset across all dimensions, resulting in an averaged centre denoted as $C$. Secondly, as described by Kurita (2020), an optimal regression line passing through $C$ is determined while minimizing the sum of the squares of projection costs in Euclidean space. Considering the distances between $C$ and points in the SWE training set $x_i$ as fixed, the minimization is achieved by maxing the squares of the scores sum. Consequently, the first principal component $w_{(1)}$ is obtained by the following, where $w$ is the weight vector.

$$w_{(1)} = argmax\{\frac{w^T X^T X w}{w^T w}\}$$

Recursively, the remaining successive principal components are derived similarly but constrained to be orthogonal to all preceding components. These eigenvectors are sorted in descending order based on their corresponding eigenvalues.

$$w_{k+1} \cdot w_1 = 0$$

$$w_{k+1} \cdot w_2 = 0$$

$$\vdots$$

$$w_{k+1} \cdot w_k = 0$$

Finally, the reduced-dimensional data is obtained by projecting the original SWE data onto the new coordinate system formed by required eigenvectors.

### 2.3 Long Short-term Memory (LSTM)

As a distinctive recurrent neural network, LSTM is designed to capture the long-term dependencies in time series data (Chang et al., 2019), which mitigates issues like gradient explosion. Hochreiter and Schmidhuber (2021) propose that the central principle of LSTM is its cell, which operates along a repetitive module chain of neural networks with subtle linear communications. This special architecture contributes to the propagation and retention of information within networks.

Three gates play crucial roles in regulating the cell to selectively retain or discard information in LSTM. They are processed by fully connected layers with sigmoid activation functions, denoted as $\sigma$.

Specifically, the Forget gate $f_t$ determines the extent to which previous information should be forgotten based on the current input $x_t$ and the hidden state $h_{t-1}$ from the previous time step, which is mathematically expressed as $f_t = \sigma(W_{xf}x_t + W_{hf}h_{t-1} + b_f)$, where $W$ and $b$ are respectively the weight and bias parameters. Furthermore, the Input gate $i_t$ relies on the calculation, $i_t = \sigma(W_{xi}x_t + W_{hi}h_{t-1} + b_i)$, to assess incorporated information to the cell state.

A candidate cell $\tilde{C}_t$ is introduced to hold the updated information, which is computed with the hyperbolic tangent activation as $\tilde{C}_t = \tanh(W_{xc}x_t + W_{hc}h_{t-1} + b_c)$. Next, the cell state is iterated by:

$$C_t = f_t C_{t-1} + i_t \tilde{C}_t$$

Finally, information is determined to flow out by the Output gate, $o_t = \sigma(W_{xo}x_t + W_{ho}h_{t-1} + b_o)$ and the current hidden status $h_t = o_t \tanh(C_t)$ is prepared for the next update.

The whole "u" testing dataset is utilized as input for making forecasts. Predicting the entire dataset collectively improves the computational efficiency compared to rollout steps, providing precise representations of the overall LSTM performance.

## 2.4 Random Forest (RF)

Initially proposed by Breiman (2001), RF represents an ensemble learning technique used for classification, regression, and other relevant tasks. This method involves constructing multiple decision trees during the training phase. According to Chutia et al. (2017), RF is characterized as a classifier comprising a collection of tree-structured classifiers,

$$\{h(x, \Theta k), k = 1, \dots\}$$

where $\Theta k$ denotes the independent identically distributed random vectors, and each tree assigns a unit vote to the most prevalent class for a given input $x$.

RF proceeds "u" data through the following stages. Firstly, a bootstrapped sample is generated by randomly drawing from the "u" training dataset with replacement. This process facilitates the creation of numerous decision trees, constructed from distinct bootstrapped training samples, with some trees possibly absent from specific sets.

Subsequently, at each decision split within a tree, a random subset of $m$ predictors is selected from the complete set of $p$ predictors to serve as candidates for the split. Only one of these $m$ predictors is then utilized for the actual split.

Moreover, the chosen of a fresh subset of $m$ predictors takes place at each split, with $m$ typically chosen to be approximately equal to the square root of $p$ (i.e., $m \approx \sqrt{p}$). During the prediction phase, each tree in the forest generates a target value prediction for a given input. In the case of this regression task on "u" predictions, results are derived from averaging the outputs of all the trees.

## 2.5 Polynomial Regression (PR)

PR constitutes a regression analysis that utilizes basis functions to model the relationship between two variables, aiming to minimize the variance of unbiased estimators.

After the data pre-processing on SWE "u", the least squares method, whose foundational principles are elucidated by Ostertagová (2012), is employed to determine the regression coefficients $b_j$ associated with each polynomial term $\varepsilon_i$, ensuring an optimal fit for the regression curve. Mathematically, for each prediction $y_i$ and the corresponding true data $y_i'$

$$y_i = \sum_{j=0}^{m} b_j x_i^j + \varepsilon_i = b_0 + b_1 x_i + b_2 x_i^2 + \cdots + b_m x_m^m + \varepsilon_i$$

which is provided by minimizing $\sum_{j=1}^{n} \varepsilon_i = \sum_{j=1}^{n} (y_i' - y_i)^2$.

In this study, the whole "u" testing dataset is input into PR. Next, PR finds the best-fitting polynomial function by minimizing the residual sum of squares, which benefits to make predictions approaching the actual "u" value.

## 2.6 Kalman Filter (KF)

KF is a recursive state estimation algorithm based on Bayesian filtering theory, which is broadly employed in linear dynamic systems with noisy measurement (Prakash et al., 2010). The optimal inference is found by combining the forecast information with actual observations to improve the accuracy of predictions.

Indeed, we take predicted results of "u" latent space as the prediction value of the current state $\hat{x}'_k$ in this study. With the covariance matrix $P'_k$ between the truth and forecasts, the Kalman gain $K_k$ is computed for the state update. The specific calculation, referring to Li et al. (2015), is shown as follows:

$$\hat{x}'_k = A\,\hat{x}'_{k-1} + B\,u_{k-1}$$

$$P'_k = A\,P_{k-1}\,A^T + Q$$

$$K_k = P'_k\,H^T\,(H\,P'_k\,H^T + R)^{-1}$$

In the above, $A$ is the state transition matrix, $B$ and $u_{k-1}$ respectively stands for the control input matrix and the control input, $Q$ and $R$ individually represents the covariance matrix of noise in the process and the observation, $H$ is marked as the observation matrix.

Then, the state is updated by the observed values $z_k$ and $K_k$ while the uncertainty of the state estimation is adjusted through the updated covariance matrix.

$$x_k = x'_k + K_k\,(z_k - H x'_k)$$

$$P_k = (I - K_k\,H)\,P'_k$$

With these continuously updating steps, KF can gradually optimize the estimation of the state, which improves the prediction performances and reduces the influence of noise.

## 3. Results Analysis

This section demonstrates the numerical results, visualizations and performance evaluations of the approaches previously introduced. Through the comparative analysis and interpretation of these results, we can choose the appropriate method for the SWE dataset to improve the accuracy and reliability of predictions.

### 3.1 PCA and CAE Comparisons

In the initial phase, we calculate the cumulative explained variance ratios (CEVR) of PCA with various dimensions to select the appropriate dimension of the latent space. This process involves striking a balance between CEVR and the number of dimensions to effectively retain essential features of SWE data. Generally, a suitable dimension level is chosen where the corresponding CEVR value surpasses 95%, following the approach outlined by Jolliffe and Cadima (2016). The result is presented afterwards.

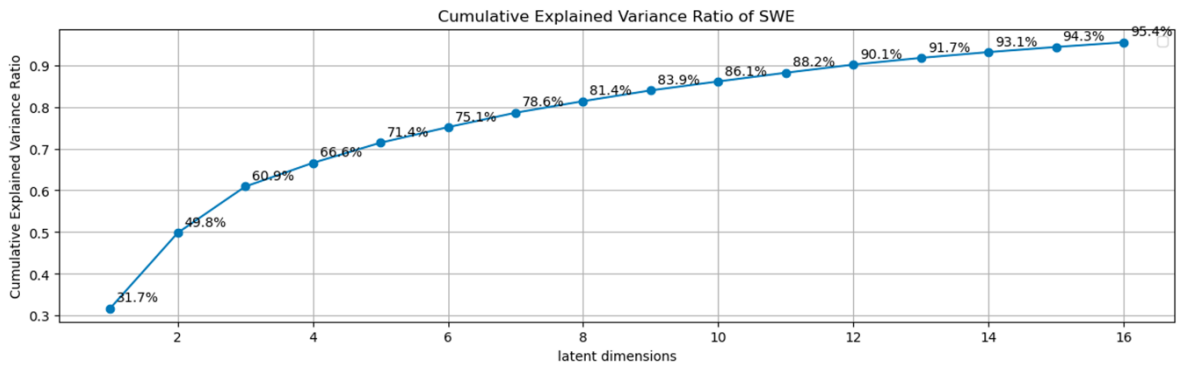

Figure1. CEVR of SWE data with different dimensions in PCA

Figure 1 depicts that when the dimension of latent space reaches 16, its CEVR achieves 95.4%, which is larger than the predetermined threshold value of 95%. Consequently, the latent space of 16 dimensions is considered a reasonable selection for SWE data in the implementations of both PCA and CAE.

Then, we proceed to investigate the reconstructive capabilities of PCA and CAE methods concerning SWE data. Regarding the evaluation of data estimation accuracy, mean square error (MSE) stands as a common metric while the lower MSE indicates the reduced bias (Hayatbini et al., 2019). Specifically, we compute MSE values for both approaches and present the results in the subsequent table.

| Methods | u | v | h |
|---|---|---|---|
| PCA | 0.0045302190431986336 | 0.004530219043198634 | 0.41981382340078816 |
| CAE | 0.28903577406580017 | 0.28789266287370324 | 0.41986656383045684 |

Table1. MSE values of PCA, CAE of u, v and h data in SWE

Table 1 delineates the statistical analysis in MSE values of the "u", "v" and "h" data after employing PCA and CAE methods on SWE. The comparison of the first two columns reveals a large discrepancy in MSE values of the "u" and "v" data between the two methods: PCA produces a value, which is approximately 0.00045 while the result of CAE is about 0.29. This apparent contrast implies that PCA is more accurate than CAE in reconstructing the data, effectively preserving more essential features of the original "u" and "v" datasets. In terms of CAE, there may be certain fluctuations in the reconstruction process. One of the reasons is that the Maxpooling layers may overamplify the dominating features and ignore some subtle details during the feature extraction. Additionally, MSE values of the "h" data in the two methods have no significant difference, both are roughly 0.42. Therefore, based on these numerical results, it remains inconclusive as to which method performs better on the reconstruction of "h" data. In the following, we show the visualization of original SWE data and reconstruction results of two methods to make an intuitive comparison and explore further.

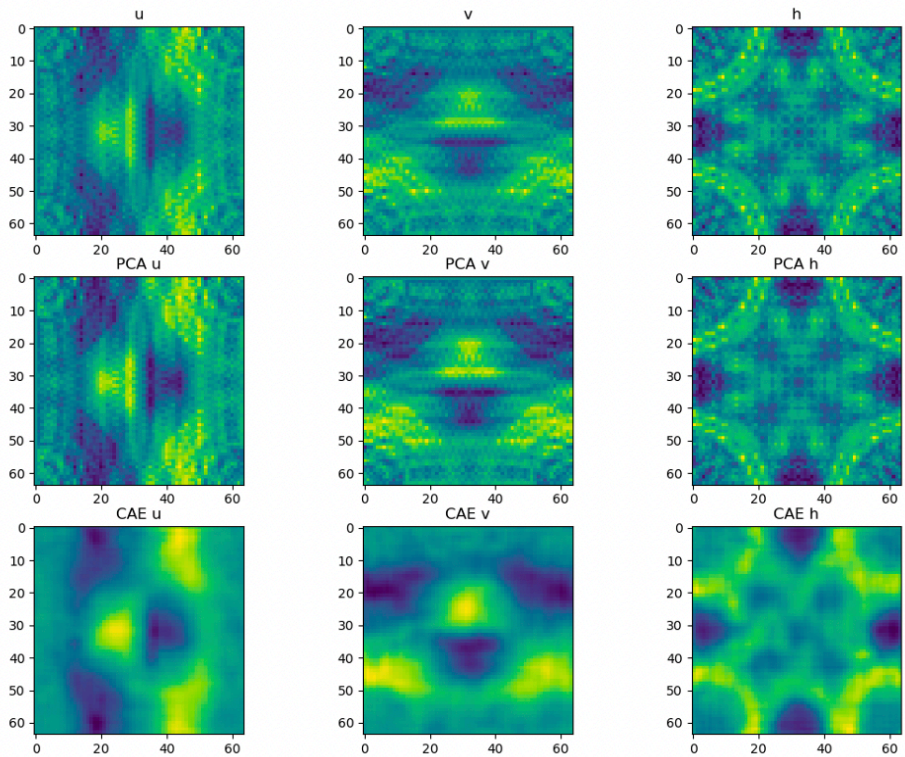

Figure2. SWE reconstructions in CAE and PCA, where "u" and "v" represent the horizontal and vertical velocity components of the fluid flow, while "h" denotes either the water surface height or the water depth relative to the reference water level.

In Figure 2, the top row displays three reference graphs which represent the actual cases. With the observation of PCA and CAE results in the subsequent graphs, it is evident that PCA yields significantly closer approximations to the reference graphs from all perspectives. This finding suggests that PCA excels in the task of data reconstruction in SWE. Specifically, upon conducting a vertical comparison in the final column of Figure 2, the differences between PCA and CAE in the "h" data images become apparent. It is acknowledged that CAE-h image can capture the various in the height of the water surface or the depth, effectively reflecting the primary characteristics of the "h" data. Nevertheless, it suffers from substantial loss of details in the visual depiction and this deficiency is particularly noticeable in some local features. For example, PCA demonstrates superior accuracy in reconstructing subtle interactions of water waves while CAE exhibits suboptimal performances in these areas. Furthermore, compared with the PCA-h image, the visual articulation of CAEs is memorably inferior with displayed edges and textures appearing considerably more blurred.

In summary, the results indicate that PCA outperforms CAE in the context of reconstructing SWE data. Therefore, further study and analysis will focus on the PCA results in the following research.

### 3.2 Three Predictive Models

Given the highly analogous structures observed between the "u", "v" and "h" data, our primary attention is focused on "u" as the target variable for subsequent SWE predictions. At first, we partition the "u" data into training and testing sets with a ratio of 8:2 and apply PCA to reduce the feature dimensions. Next, the complete testing dataset is utilized as input for anticipating behaviours of the "u" data in the latent space, employing three distinct forecasting models: Long Short-Term Memory (LSTM), Random Forest (RF), and Polynomial Regression (PR). The ensuing figure illustrates the comparisons between the prognosticated curve for each method with the corresponding actual value.

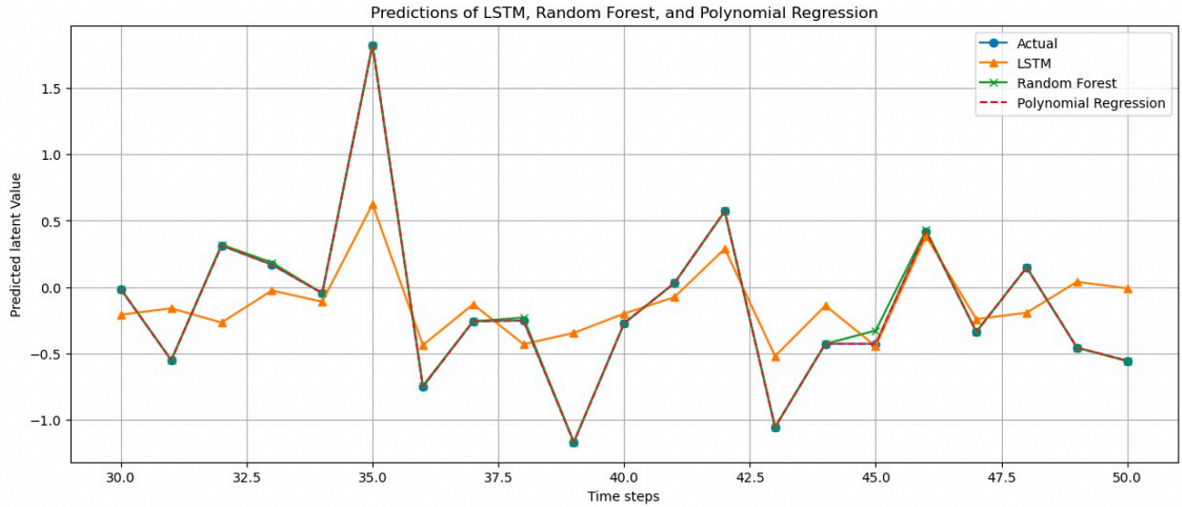

Figure3. Contrast the actual values with the predicted values obtained through the utilization of LSTM, RF, and PR techniques, respectively.

Figure 3 presents the subsets with time steps ranging from 30 to 50 units of local predictions truncated from a total of 1000 testing results, which facilitates our comparative analysis. As for the original data trend, its values primarily fluctuate between -1 and 0.5 units. However, at the 35-time step, the actual value of "u" in the latent space substantially increases to approximately 1.8 units and is considered a relatively unstable point within this local interval, potentially posing challenges for forecasts.

Upon applying LSTM, its forecast trend exhibits similarity to the origin data. Nevertheless, there are considerable discrepancies at time steps 35 and 39, with deviations of about 1 and 0.8 units, respectively, while the others are generally below 0.5 units. This indicates that LSTM may have poor performance in anticipating abrupt value changes.

In contrast, RF yields a relatively stable predictive curve, which is closely aligned with the actual data, with only slight differences observed at specific points. For instance, at the time step of 45, the prognostication has a relatively significant increase, which may be attributed to the overfitting of RF in the process.

Finally, the red dashed line in Figure 3 comes closest to the original trend, signifying that PR is superior at forecasting latent behaviours of the "u" data.

Apart from the prediction graphs, comparing MSE and $R^2$ values of each method is also available for the evaluation of their anticipated capability. Schneider et al. (2010) set forth that $R^2$ measures the prognostication explanation on the overall variance and its value closer to 1 indicates better forecast performance. The table below describes the computed MSE and $R^2$ values of all three predictive models.

| Methods | LSTM | Random Forest | Polynomial Regression |
|---|---|---|---|
| MSE | 0.4656647741794685 | 0.00015462584800998136 | 5.636266193412578e-29 |
| $R^2$ | 0.8057731106736332 | 0.9998672270557152 | 1.0 |

Table2. MSE and $R^2$ values of LSTM, RF and PR methods

Based on the numerical findings in Table 2, three models demonstrate varying accuracy and suitability in predicting "u" data within the latent space. Firstly, we examine values across three columns and find that a higher value of $R^2$ corresponds to a relatively smaller MSE, which conforms to their negative correlation. Subsequently, a row-wise analysis of Table 2 facilitates the comparative assessment of forecast performances among the three models.

PR showcases strong predictive capabilities with its MSE approaching 0 and almost negligible, which implies the subtle discrepancies between the anticipations and true data. Meanwhile, the perfect $R^2$ value of 1.0 suggests an exceptionally elevated level of forecast accuracy.

Similarly, RF also makes excellent prognostications on the "u" data, as evidenced by its MSE value of 0.00015 and $R^2$ value of 0.9999. Nonetheless, in comparison to the above method, RF slightly lags in performance.

On the other hand, although LSTM shows its ability to capture the data trend, 0.466 MSE is larger than the other two values while 0.81 $R^2$ is considerably lower, indicating that predictions are significantly different from the origin data at specific time steps.

As a supplement, we project the anticipated latent data back into the full "u" space through the inverse transformation of PCA to generate the forecast "u" images. With the observation of details revealed in the images, it becomes plausible to compare the prediction accuracy of the three models intuitively and visually.

To evaluate image quality and compare their similarities, the Structural Similarity Index (SSIM) and Peak Signal-to-Noise Ratio (PSNR) are usual indicators (Horé and Ziou, 2010).

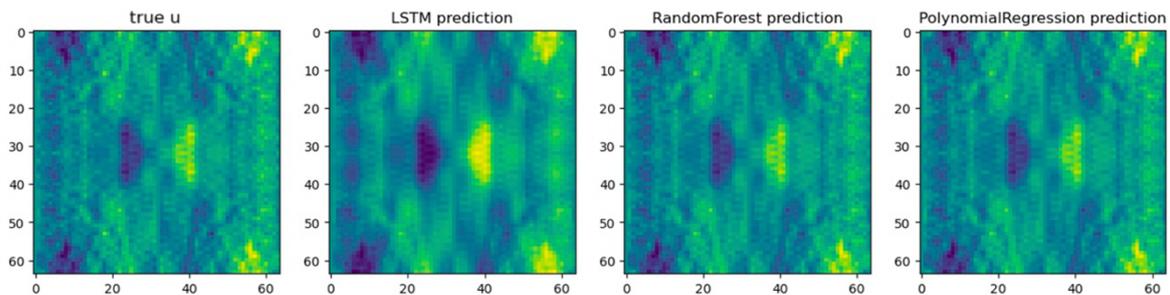

| Methods | LSTM | Random Forest | Polynomial Regression |
|---|---|---|---|
| SSIM | 0.8461 | 0.9881 | 0.9882 |
| PSNR | 33.2363 | 42.9684 | 42.9938 |

Figure4. The visualization of actual "u" and three predictive results in the full space, showing their SSIM and PSNR values of three forecast images, compared to the first one.

Through a meticulous examination of four images in Figure 4, we notice a concurrence among predictive graphs of RF and PR with the actual "u" data. Their SSIM values are roughly 0.988, which suggests that the commonality between the first image and these two images lies in the remarkable structural and content similarity. Indeed, superior image quality is indicated in PR with a larger PSNR value of about 42.99 than RF's 42.97 at around.

However, with smaller SSIM and PSNR values of 0.8461 and 33.24, LSTM presents a relatively ambiguous image, especially several details missing in the representation of some subtle features. Additionally, LSTM prioritizes capturing dominant features of the "u" data, with dark blocks appearing darker and light blocks appearing lighter in the second graph, which implies that LSTM tends to amplify extreme values when prognosticating the horizontal velocity of water flow.

Conclusively, as for the SWE "u" data, PR has an optimal performance on the forecasts, along with robustness and efficiency. And predictions of RF closely approximate the actual "u" values, underscoring its anticipated capability. However, LSTM results fall short in comparison, which suggests the requirement for additional optimization and adjustments to improve its accuracy in forecasting finer details.

### 3.3 Data Assimilation

Within this part, we employ the Kalman filter (KF) as a data assimilation method to improve the accuracy of model prognostications, which is achieved by establishing a correlation between forecast outcomes and actual "u" values. Following the KF implementation, we re-evaluate MSE and $R^2$ values of three predictive models. Then, a comparative analysis is conducted, comparing these new results with the previous ones, to assess the impact of the Kalman filter on the model anticipations.

| Methods | LSTM with KF | Random Forest with KF | Polynomial Regression with KF |
|---|---|---|---|
| MSE | 0.0004695465584015516 | 0.0006581351248812933 | 0.0004695465584015541 |
| $R^2$ | 0.9996872284124825 | 0.9995399057528679 | 0.9996872284124825 |

Table3. MSE and $R^2$ values of LSTM, RF and PR methods after the implementation of KF

By observing the last two columns of Table 3, it becomes evident that for both RF and PF, there are negligible differences between MSE and $R^2$ values after applying KF and those obtained without. This finding demonstrates that KF does not significantly contribute to improving the forecast accuracy of RF and PF. Indeed, it may even disrupt the prediction process for these models.

However, when KF is employed in LSTM, a remarkable enhancement in the anticipation performance is observed. Specifically, the MSE and $R^2$ values of LSTM reach 0.00089 and 0.9998, respectively, indicating a substantial improvement in predictive capabilities. These results establish LSTM with the implementation of KF as a competent and proficient model for accurate representation in this context.

Additionally, the extent of improvement in the accuracy of the LSTM model achieved through the integration of KF is elucidated in the following. To ensure a lucid comparison, we present the forecast curves and images, both with and without the application of KF.

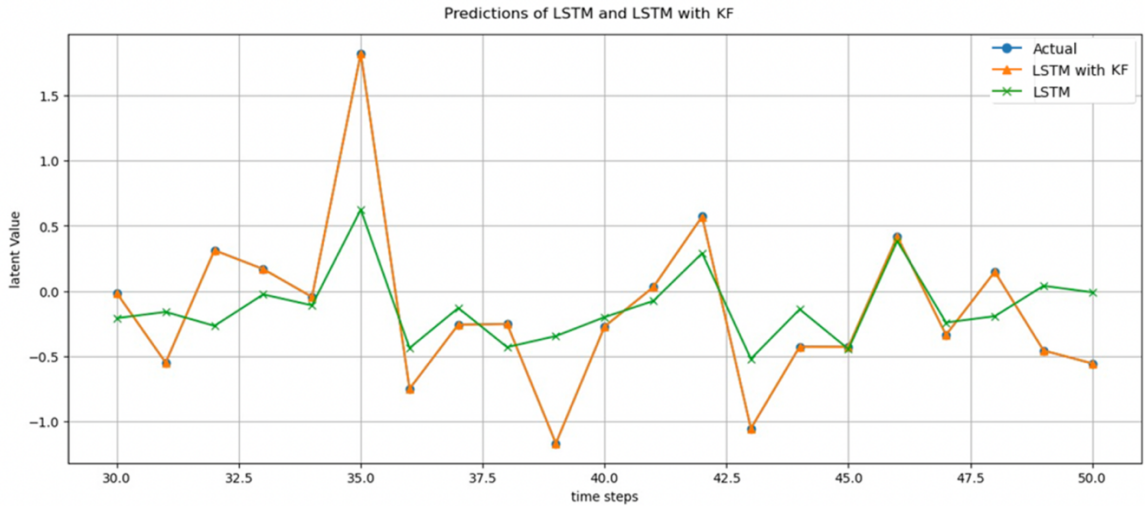

Figure5. the actual "u" data in the latent space, prediction curves of LSTM and LSTM with the application of KF

In Figure 5, the yellow and blue areas reveal a high overlap, depicting that the prediction result of LSTM through KF is highly analogous to the true latent "u" data.

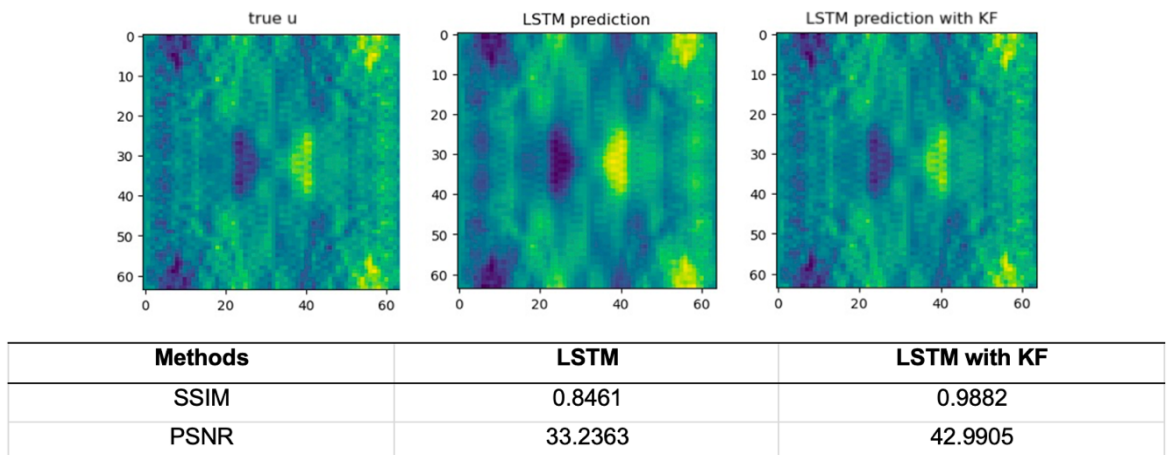

| Methods | LSTM | LSTM with KF |
|---------|------|--------------|
| SSIM | 0.8461 | 0.9882 |
| PSNR | 33.2363 | 42.9905 |

Figure6. The visualization of true "u" data, two predictive images of LSTM with and without KF, coupled with their SSIM and PSNR values compared to the true one.

At the same time, the comparison of the latter two figures in Figure 6 illustrates that the LSTM prognostication after KF contains some lost details and its clarity is better, which also proves the improved accuracy of forecasts. Meanwhile, numerical comparisons show that with the implementation of KF on LSTM, its SSIM and PSNR values are approximately 0.99 and 42.99, correcting the noise and detail differences in the image and improving the similarity to the real image.

To conclude, the application of KF exhibits notable advantages in enhancing LSTM predictive capabilities for the "u" data in the latent space. Nevertheless, KF seems to have no influence on the prediction accuracy of RF and PR approaches based on this dataset.

## 4. Conclusions and Discussions

In summary, a comprehensive methodology is proposed to make accurate forecasts of SWE data. The study employs three distinct predictive models and evaluates their performances in anticipating the SWE potential behaviours, with the incorporation of ROM and DA techniques. Initially, SWE data undergoes dimensionality reduction via CAE and PCA, projected into the latent space. Comparative analysis of MSE values and image visualizations indicates that PCA effectively captures more essential features in the SWE reconstruction, highlighting its superior accuracy compared to CAE. Then, LSTM, RF and PR models are utilized to forecast the latent spatial SWE data, assessed by numerical metrics, and restored data visualizations in the full space. Results suggest that PR exhibits the best performance, RF is the second while LSTM demonstrates relatively lower accuracy in predictions, especially concerning nuanced features. Nevertheless, KF is implemented to substantially enhance the accuracy of LSTM and its impact on RF and PR remains negligible in the context of SWE data. Ultimately, the culmination of these efforts is reflected in Figure 7, where three methods collectively achieve precise forecasts.

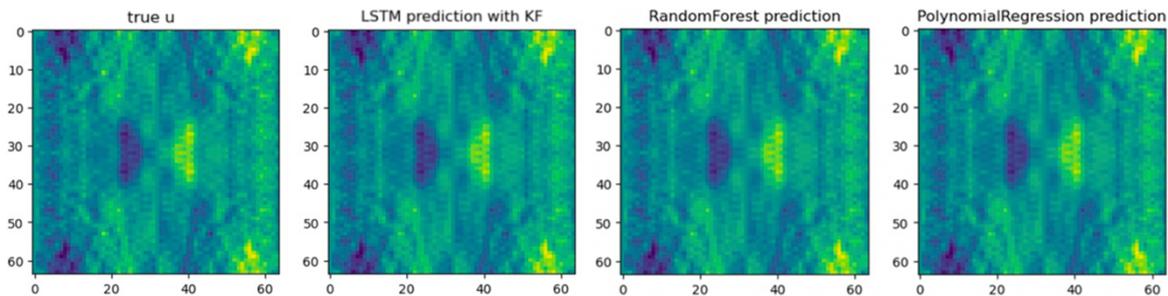

Figure7. the actual "u" image and accurate predictions of LSTM, RF and PR methods

This investigation combines abundant techniques to improve the prognostications accuracy of SWE, which holds significant implications for the application in anticipating storms and tsunamis in natural disaster management. However, Tort et al. (2014) propose that given the near-spherical morphology of the Earth, the shallow water phenomenon on the surface has a pronounced spatial correlation on a global scale. Concurrently, the planetary rotation and axial tilt engender diurnal and seasonal variations across the spatial distribution of SWE, constituting additional challenges in the forecasts. Indeed, the discontinuous Galerkin method with wet/dry transitions can establish a harmonious balance on the spherical SWE, which is suitable for predicting large-scale tsunami events (Bonev et al., 2018). Furthermore, Lanser et al. (2000) leverage the Osher's finite-volume scheme to resolve the spatial discretization of SWE within the spherical space. With these complexities, a more profound exploration of numerical simulations in SWE predictions stands as a promising avenue for future research endeavours. Based on the variability of terrain and other environmental impacts, the forecasting model should be continuously improved to adapt to different and complex requirements.

## 5. References

Bach, E. and Ghil, M. (2023) 'A multi-model ensemble Kalman filter for data assimilation and forecasting', Journal of Advances in Modeling Earth Systems, 15(1). doi:10.1029/2022ms003123.

Bonev, B. et al. (2018) 'Discontinuous galerkin scheme for the spherical shallow water equations with applications to tsunami modeling and prediction', Journal of Computational Physics, 362, pp. 425–448. doi: 10.1016/j.jcp.2018.02.008.

Breiman, L. (2001) Machine Learning, 45(1), pp. 5–32. doi:10.1023/a:1010933404324.

Cavaleri, L. and Rizzoli, P.M. (1981) 'Wind wave prediction in shallow water: Theory and applications', Journal of Geophysical Research, 86(C11), pp. 10961–10973. doi:10.1029/jc086ic11p10961.

Chang, B. et al. (2019) AntisymmetricRNN: A dynamical system view on recurrent neural networks, arXiv.org. Available at: https://doi.org/10.48550/arXiv.1902.09689 (Accessed: 10 August 2023).

Cheng, S. et al. (2023) 'Machine learning with data assimilation and uncertainty quantification for dynamical systems: A Review', IEEE/CAA Journal of Automatica Sinica, 10(6), pp. 1361–1387. doi: 10.1109/jas.2023.123537.

Cheng, S., Chen, J., Anastasiou, C. et al. (2023) Generalised Latent Assimilation in Heterogeneous Reduced Spaces with Machine Learning Surrogate Models. J Sci Comput 94, 11. doi:10.1007/s10915-022-02059-4.

Cheng, S., Prentice, I.C., et al. (2022) 'Data-driven surrogate model with Latent Data Assimilation: Application to wildfire forecasting', Journal of Computational Physics, 464, p. 111302. doi: 10.1016/j.jcp.2022.111302.

Chutia, D. et al. (2017) 'An effective ensemble classification framework using random forests and a correlation based feature selection technique', Transactions in GIS, 21(6), pp. 1165–1178. doi:10.1111/tgis.12268.

Dellar, P.J. and Salmon, R. (2005) 'Shallow water equations with a complete coriolis force and topography', Physics of Fluids, 17(10). doi:10.1063/1.2116747.

Ding, S. et al. (2013) 'Extreme learning machine: Algorithm, Theory and Applications', Artificial Intelligence Review, 44(1), pp. 103–115. doi:10.1007/s10462-013-9405-z.

Dinther, Y.V., Künsch, H. and Fichtner, A. (2019) 'Ensemble data assimilation for earthquake sequences: Probabilistic estimation and forecasting of fault stresses', Geophysical Journal International, 217(3), pp. 1453–1478. doi:10.1093/gji/ggz063.

Franca, M.J., Valero, D. and Liu, X. (2022) 'Turbulence and rivers', Treatise on Geomorphology, 6.1, pp. 151–175. doi:10.1016/b978-0-12-818234-5.00135-8.

Gong, H. et al. (2022) 'An efficient digital twin based on machine learning SVD autoencoder and Generalised Latent Assimilation for Nuclear Reactor Physics', Annals of Nuclear Energy, 179, p. 109431. doi: 10.1016/j.anucene.2022.109431.

Han, J., Kamber, M. and Pei, J. (2012) 'Outlier detection', Data Mining, pp. 543–584. doi:10.1016/b978-0-12-381479-1.00012-5.

Hayatbini, N. et al. (2019) 'Conditional generative adversarial networks (cgans) for near real-time precipitation estimation from multispectral GOES-16 satellite imageries—PERSIANN-cgan', Remote Sensing, 11(19), p. 2193. doi:10.3390/rs11192193.

Hendricks, E.A., Vigh, J.L. and Rozoff, C.M. (2021) 'Forced, balanced, axisymmetric shallow water model for understanding short-term tropical cyclone intensity and wind structure changes', Atmosphere, 12(10), p. 1308. doi:10.3390/atmos12101308.

Hernandez, B. et al. (2023) 'Learning meaningful latent space representations for patient risk stratification: Model Development and validation for dengue and other acute febrile illness', Frontiers in Digital Health, 5. doi:10.3389/fdgth.2023.1057467.

Hotelling, H. (1933) 'Analysis of a complex of statistical variables into principal components.', Journal of Educational Psychology, 24(6), pp. 417–441. doi:10.1037/h0071325.

Hochreiter, S. and Schmidhuber, J. (2021) LSTM can solve hard long time lag problems - neurips, *Advances in Neural Information Processing Systems 9 (NIPS 1996)*, pp. 474-479. Available at: https://proceedings.neurips.cc/paper/1996/file/a4d2f0d23dcc84ce983ff9157f8b7f88-Paper.pdf (Accessed: 10 August 2023).

Horé, A. and Ziou, D. 'Image Quality Metrics: PSNR vs. SSIM,' 2010 20th International Conference on Pattern Recognition, Istanbul, Turkey, 2010, pp. 2366-2369, doi: 10.1109/ICPR.2010.579.

Idier, D. et al. (2019) 'Interactions between mean sea level, tide, surge, waves and flooding: Mechanisms and contributions to sea level variations at the Coast', Surveys in Geophysics, 40(6), pp. 1603–1630. doi:10.1007/s10712-019-09549-5.

Intergovernmental Panel on Climate Change (IPCC) (2022) "Sea Level Rise and Implications for Low-Lying Islands, Coasts and Communities," in The Ocean and Cryosphere in a Changing Climate: Special Report of the Intergovernmental Panel on Climate Change. Cambridge: Cambridge University Press, pp. 321–446. doi: 10.1017/9781009157964.006.

Jia, W. et al. (2022) 'Feature dimensionality reduction: A Review', Complex & Intelligent Systems, 8(3), pp. 2663–2693. doi:10.1007/s40747-021-00637-x.

Jolliffe, I.T. and Cadima, J. (2016) 'Principal component analysis: A review and recent developments', Philosophical Transactions of the Royal Society A: Mathematical, Physical and Engineering Sciences, 374(2065), p. 20150202. doi:10.1098/rsta.2015.0202.

Kadeethum, T. et al. (2022) 'Enhancing high-fidelity nonlinear solver with reduced order model', Scientific Reports, 12(1). doi:10.1038/s41598-022-22407-6.

Karasözen, B., Yıldız, S. and Uzunca, M. (2020) 'Structure preserving model order reduction of shallow water equations', Mathematical Methods in the Applied Sciences, 44(1), pp. 476–492. doi:10.1002/mma.6751.

Khalaf, M. et al. (2020) 'IOT-enabled Flood severity prediction via ensemble machine learning models', IEEE Access, 8, pp. 70375–70386. doi:10.1109/access.2020.2986090.

Kim, B. et al. (2023) 'Deep neural network-based reduced-order modeling of ion–surface interactions combined with molecular dynamics simulation', Journal of Physics D: Applied Physics, 56(38), p. 384005. doi:10.1088/1361-6463/acdd7f.

Kurita, T. (2020) 'Principal Component Analysis (PCA)', Computer Vision, pp. 1–4. doi:10.1007/978-3-030-03243-2_649-1.

Lanser, D., Blom, J.G. and Verwer, J.G. (2000) 'Spatial discretization of the shallow water equations in spherical geometry using Osher's scheme', Journal of Computational Physics, 165(2), pp. 542–565. doi:10.1006/jcph.2000.6632.

Le, X. et al. (2019) 'Application of long short-term memory (LSTM) neural network for flood forecasting', Water, 11(7), p. 1387. doi:10.3390/w11071387.

Moye, M.J. and Diekman, C.O. (2018) 'Data assimilation methods for neuronal state and parameter estimation', The Journal of Mathematical Neuroscience, 8(1). doi:10.1186/s13408-018-0066-8.

Ostertagová, E. (2012) 'Modelling using polynomial regression', Procedia Engineering, 48, pp. 500–506. doi: 10.1016/j.proeng.2012.09.545.

Pearson, K. (1901) 'On lines and planes of closest fit to systems of points in space', The London, Edinburgh, and Dublin Philosophical Magazine and Journal of Science, 2(11), pp. 559–572. doi:10.1080/14786440109462720.

Pintelas, E. and Pintelas, P. (2022) 'A 3D-CAE-CNN model for deep representation learning of 3D images', Engineering Applications of Artificial Intelligence, 113, p. 104978. doi: 10.1016/j.engappai.2022.104978.

Pinto-Ledezma, J.N. and Cavender-Bares, J. (2021) 'Predicting species distributions and community composition using satellite remote sensing predictors', Scientific Reports, 11(1). doi:10.1038/s41598-021-96047-7.

Prakash, J., Patwardhan, S.C. and Shah, S.L. (2010) 'Constrained Nonlinear State estimation using ensemble Kalman Filters', *Industrial and Engineering Chemistry Research*, 49(5), pp. 2242–2253. doi:10.1021/ie900197s.

Li, Q. et al. (2015) 'Kalman filter and its Application', 2015 8th International Conference on Intelligent Networks and Intelligent Systems (ICINIS). doi:10.1109/icinis.2015.35.

Rocchi, L., Chiari, L. and Cappello, A. (2004) 'Feature selection of stabilometric parameters based on principal component analysis', Medical and Biological Engineering and Computing, 42(1), pp. 71–79. doi:10.1007/bf02351013.

Sarkar, S., Biswas, R. and Ghosh, A.K. (2019) 'On some graph-based two-sample tests for high dimension, low sample size data', Machine Learning, 109(2), pp. 279–306. doi:10.1007/s10994-019-05857-4.

Schneider, A., Hommel, G. and Blettner, M. (2010) 'Linear regression analysis', Deutsches Ärzteblatt international. doi:10.3238/arztebl.2010.0776.

Scheinost, D. et al. (2019) 'Ten simple rules for predictive modeling of individual differences in neuroimaging', NeuroImage, 193, pp. 35–45. doi: 10.1016/j.neuroimage.2019.02.057.

Tort, M. et al. (2014) 'Consistent shallow-water equations on the rotating sphere with complete coriolis force and topography', Journal of Fluid Mechanics, 748, pp. 789–821. doi:10.1017/jfm.2014.172.

Tripathi, S. et al. (2021) 'Ensuring the robustness and reliability of data-driven knowledge discovery models in production and manufacturing', Frontiers in Artificial Intelligence, 4. doi:10.3389/frai.2021.576892.

Zhai, S. et al. (2017) 'S3pool: Pooling with Stochastic Spatial sampling', 2017 IEEE Conference on Computer Vision and Pattern Recognition (CVPR). doi:10.1109/cvpr.2017.426.

Zhong, C. et al. (2023) 'Reduced-order digital twin and latent data assimilation for global wildfire prediction', Natural Hazards and Earth System Sciences, 23(5), pp. 1755–1768. doi:10.5194/nhess-23-1755-2023.

Zhuang, Y. et al. (2022) 'Ensemble latent assimilation with deep learning surrogate model: Application to drop interaction in a microfluidics device', Lab on a Chip, 22(17), pp. 3187–3202. doi: 10.1039/d2lc00303a.